\documentclass[conference]{IEEEtran}
\IEEEoverridecommandlockouts

\usepackage{cite}
\usepackage{amsmath,amssymb,amsfonts}
\usepackage{graphicx}
\usepackage{textcomp}
\usepackage{xcolor}
\usepackage{booktabs}
\usepackage{balance}
\usepackage[printonlyused]{acronym}
\usepackage{siunitx}
\usepackage[linesnumbered,ruled,vlined]{algorithm2e}

\acrodef{fh}[FH]{frequency hopping}
\graphicspath{{figures/}}

\newcommand{\FGRu}[1]{Figure~\ref{#1}}
\newcommand{\SEC}[1]{Section~\ref{#1}}
\newcommand{\TAB}[1]{Table~\ref{#1}}
\newcommand{\Eq}[1]{Equation~\ref{#1}}

\def\BibTeX{{\rm B\kern-.05em{\sc i\kern-.025em b}\kern-.08em
    T\kern-.1667em\lower.7ex\hbox{E}\kern-.125emX}}

\begin{document}

\title{Joint Direction-of-Arrival and Range Estimation for Millimeter-Wave Uniform Linear Array Radar}

\author{\IEEEauthorblockN{Necati Kagan Erkek, Zeynep Gul Pehlivanli}
\IEEEauthorblockA{Telecommunications Engineering, Department of Electronics, Information and Bioengineering\\
Politecnico di Milano, Piazza Leonardo da Vinci 32, 20133 Milan, Italy\\
\texttt{necatikagan.erkek@mail.polimi.it, zeynepgul.pehlivanli@mail.polimi.it}}}

\maketitle

\begin{abstract}
An FFT-based direction-of-arrival (DOA) and range-estimation framework for a monostatic uniform linear array (ULA) operating at 77 GHz is presented. A narrowband sinusoidal waveform is used to derive the spatial phase model, determine an aliasing-free inter-element spacing, and select the aperture required to obtain a boresight angular resolution of 2 degree. The resulting design uses an element spacing of 0.97 mm and 58 antenna elements, corresponding to an aperture length of 56.42 mm. Numerical results show accurate angular estimation for a single target at 30 degree and for multiple simultaneous targets. The analysis is further extended to two-dimensional localization by replacing the narrowband waveform with a 1 GHz sinc-modulated signal, which provides an approximate range resolution of 0.15 m. Additional simulations quantify the effects of additive complex Gaussian noise, increased antenna spacing, and target decorrelation on the DOA response.
\end{abstract}

\begin{IEEEkeywords}
Direction of arrival, uniform linear array, mmWave radar, FFT beamforming, range estimation, radar signal processing.
\end{IEEEkeywords}

\section{Introduction}
Direction-of-arrival (DOA) estimation is a fundamental problem in array signal processing, radar sensing, wireless localization, and remote imaging. Antenna arrays exploit deterministic phase differences across spatially separated elements to infer the angular location of a radiating or reflecting source, with angular resolution governed primarily by aperture length, wavelength, and spatial sampling density \cite{balanis2016,johnson1993,van_trees2002}. Uniform linear arrays (ULAs) remain especially important because their regular geometry permits tractable signal models and efficient spatial-spectrum processing.

Classical DOA estimation techniques include conventional beam scanning, Fourier-domain spatial spectrum estimation, and high-resolution subspace methods. The MUSIC and ESPRIT algorithms represent two influential parametric approaches, while broader surveys and spectral-analysis treatments establish the associated identifiability, resolution, and noise-sensitivity trade-offs \cite{schmidt1986,roy1989,krim1996,stoica2005}. Adaptive beamforming and minimum-variance spectrum analysis provide another important line of array-processing methods, while information-theoretic criteria support model-order selection in subspace estimators \cite{capon1969,vanveen1988,wax1985,godara1997}. In practical radar systems, FFT-based processing remains attractive because it provides low implementation complexity and direct compatibility with range-Doppler processing chains \cite{richards2014,skolnik2008}. For millimeter-wave radar, the 77 GHz band is widely used in short-range sensing, where small wavelengths enable compact arrays with fine angular discrimination \cite{hasch2012}. Range-angle localization additionally requires a waveform with sufficient bandwidth, since range resolution is inversely proportional to signal bandwidth \cite{richards2014,skolnik2008,li2008}. These principles are also consistent with broader electromagnetic imaging and interferometric processing the frameworks \cite{electromagnetic_imaging,insar_principles}.

Motivated by these foundations, the reported study addresses ULA-based localization at 77 GHz under a monostatic measurement configuration. The technical objectives are threefold. First, a narrowband ULA signal model is derived from the two-way propagation delay across the array. Second, the inter-element spacing and number of elements are selected to satisfy spatial aliasing and boresight resolution constraints. Third, MATLAB simulations validate the angular estimator for single-target, multi-target, noisy, aliased, and decorrelating-target scenarios before extending the waveform model to two-dimensional range-angle estimation.

\section{System Geometry}
The considered monostatic ULA geometry is shown in \FGRu{fig:geometry}. The array contains \(N\) antenna elements displaced along the \(x\)-axis, with uniform separation \(d_x\). The figure depicts nine elements only for illustration; the designed array contains the number of elements derived in \SEC{sec:array_design}. A point target is located at range \(d\) and angular position \(\theta\) with respect to boresight.

\begin{figure}[ht!]
    \centering
    \includegraphics[width=8cm,height=5cm]{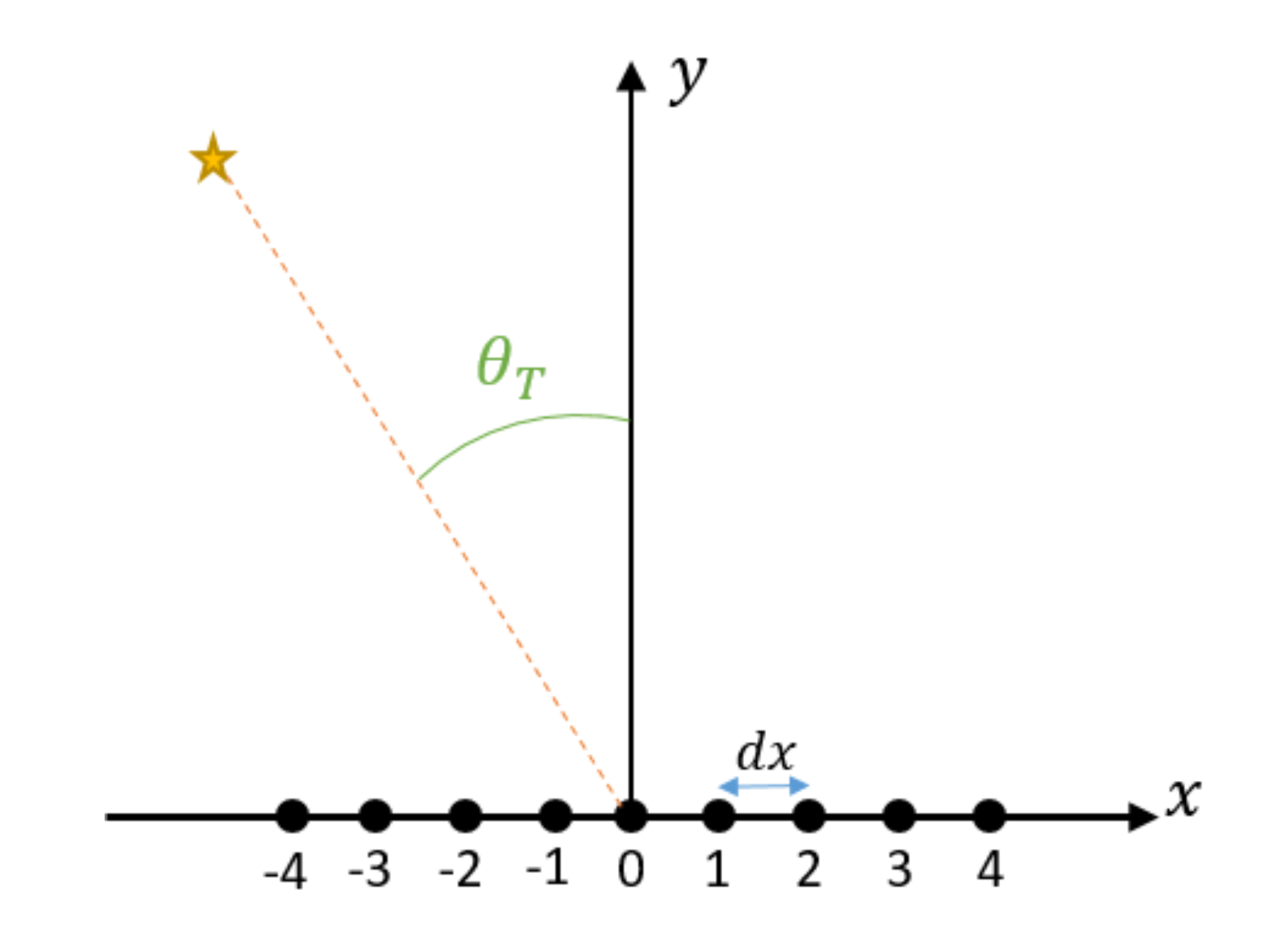}
    \caption{Monostatic ULA geometry for DOA estimation.}
    \label{fig:geometry}
\end{figure}

Each antenna transmits the narrowband signal
\begin{equation}
    g(t)=e^{j2\pi f_0 t}, \qquad f_0=77\,\mathrm{GHz},
\end{equation}
and receives its own echo. This measurement mode corresponds to a monostatic radar configuration. The far-field approximation is adopted for the angular derivation, so the path difference between adjacent elements is determined by the projected spacing along the propagation direction.

\section{Direction-of-Arrival Estimation}
\subsection{Signal Model}
For the \(n\)-th antenna element, the two-way propagation delay is
\begin{equation}
    \tau_n=\frac{2d_n}{c},
    \label{eq:delay}
\end{equation}
where \(d_n\) denotes the one-way distance from the target to the \(n\)-th element and \(c\) is the speed of light. As illustrated in \FGRu{fig:path_geometry}, the distance of the \(n\)-th element can be approximated as
\begin{equation}
    d_n=d-n d_x\sin\theta .
    \label{eq:distance}
\end{equation}
Substitution of \Eq{eq:distance} into \Eq{eq:delay} yields
\begin{equation}
    \tau_n=\frac{2(d-n d_x\sin\theta)}{c}.
    \label{eq:delay_full}
\end{equation}

\begin{figure}[ht!]
    \centering
    \includegraphics[width=7cm,height=5cm]{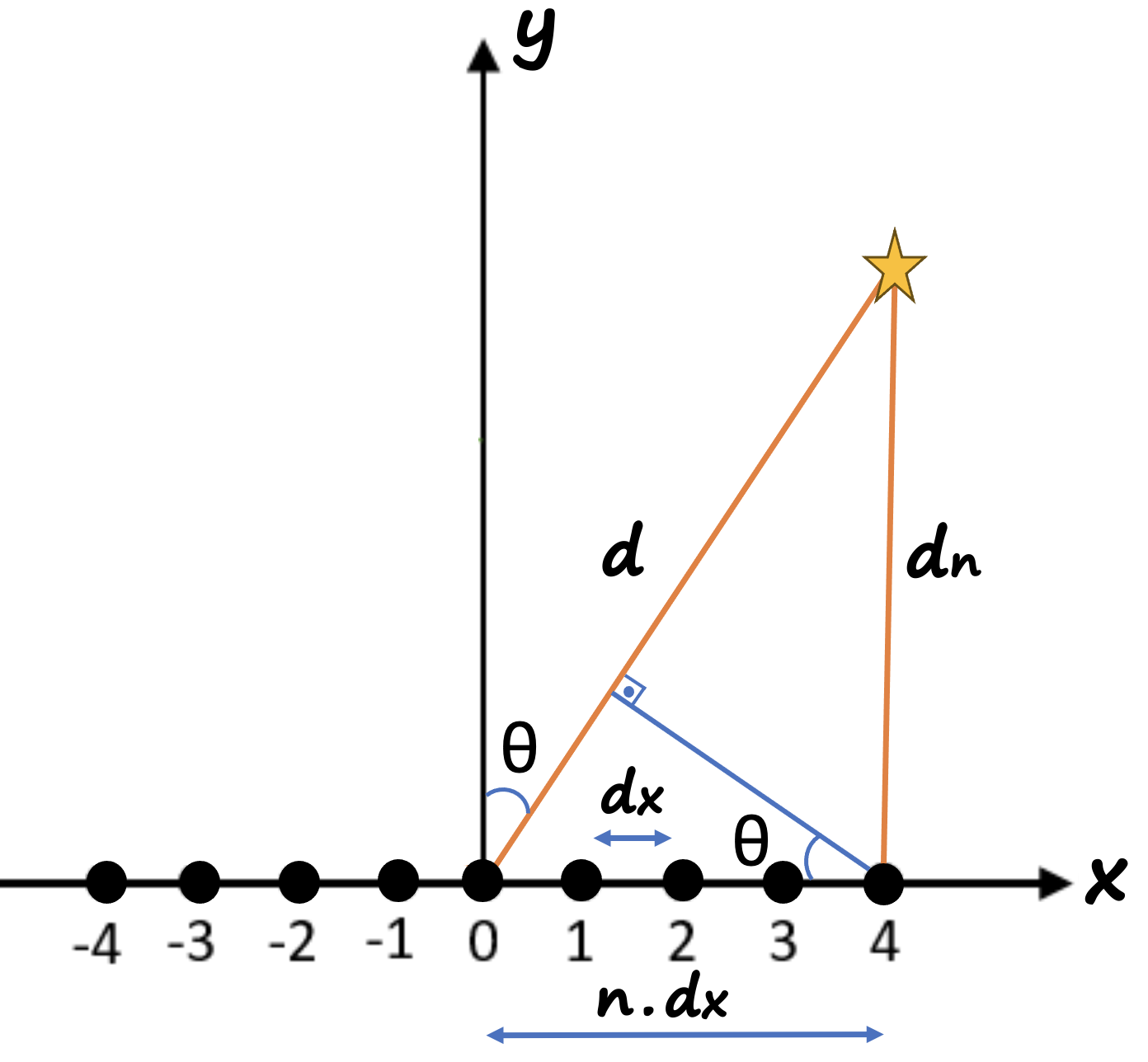}
    \caption{Path-length difference between adjacent ULA elements.}
    \label{fig:path_geometry}
\end{figure}

After ideal demodulation, the received narrowband signal at the \(n\)-th element is expressed as
\begin{equation}
\begin{aligned}
S_{Rx}(n) &= A e^{j\phi} e^{j2\pi f_0(t-\tau_n)} e^{-j2\pi f_0t} \\
          &= A e^{j\phi} e^{-j2\pi f_0 \tau_n} \\
          &= A e^{j\phi} e^{-j2\pi f_0 \frac{2(d-n d_x\sin\theta)}{c}},
\end{aligned}
\label{eq:rx_signal}
\end{equation}
where \(A\) and \(\phi\) denote the target-dependent amplitude and phase. The range-dependent term is common to all array elements and can be absorbed into
\begin{equation}
    \tilde{A}=A e^{j\phi}e^{-j2\pi f_0 \frac{2d}{c}}.
\end{equation}
The spatial response is therefore
\begin{equation}
    S_{Rx}(n)=\tilde{A}e^{j2\pi f_0\frac{2n d_x\sin\theta}{c}}.
    \label{eq:spatial_response}
\end{equation}
This expression shows that the DOA is encoded as a spatial frequency across the sampled aperture.

\subsection{Array Design and DOA Estimation}
\label{sec:array_design}
The spatial response in \Eq{eq:spatial_response} can be written as
\begin{equation}
\begin{aligned}
S_{Rx}(n) &= \tilde{A}e^{j2\pi \tilde{f}\Delta T},\\
\tilde{f} &= \frac{2f_0\sin\theta}{c},\\
\Delta T &= n d_x,
\end{aligned}
\label{eq:spatial_frequency}
\end{equation}
where \(\tilde{f}\) is the spatial frequency associated with \(\theta\). The maximum spatial frequency occurs at \(\theta=90^{\circ}\):
\begin{equation}
    \max\{\tilde{f}\}=\frac{2f_0}{c}.
    \label{eq:max_frequency}
\end{equation}
To avoid spatial aliasing in the monostatic configuration, the sampling condition is
\begin{equation}
    2\left(\frac{2f_0}{c}\right) \leq \frac{1}{d_x}.
    \label{eq:nyquist}
\end{equation}
Consequently, the selected inter-element spacing is
\begin{equation}
    d_x=\frac{c}{4f_0}=9.7276\times10^{-4}\,\mathrm{m}.
    \label{eq:spacing}
\end{equation}

The required aperture is obtained from the desired boresight resolution. For an angular separation of \(2^{\circ}\), the first-zero spacing of the finite-aperture sinc response gives
\begin{equation}
\begin{gathered}
\frac{1}{L} \leq \frac{2f_0\sin(2^{\circ})}{c}, \qquad L=N d_x,\\
N \geq \frac{c}{d_x f_0 2\sin(2^{\circ})}=57.38 .
\end{gathered}
\label{eq:aperture}
\end{equation}
The final array uses \(N=58\) elements. The resulting design parameters are summarized in \TAB{tab:design_parameters}.

\begin{table}[ht!]
\caption{ULA Design Parameters}
\centering
\begin{tabular}{lcc}
\toprule
Parameter & Symbol & Value \\
\midrule
Center frequency & \(f_0\) & 77 GHz \\
Number of elements & \(N\) & 58 \\
Element spacing & \(d_x\) & 0.97276 mm \\
Array length & \(L\) & 56.42 mm \\
\bottomrule
\end{tabular}
\label{tab:design_parameters}
\end{table}

The DOA estimate is obtained by applying an FFT to the received spatial samples and identifying the dominant spatial-frequency bin \(\hat{f}\). The angle estimate follows as
\begin{equation}
    \hat{\theta}=\arcsin\left(\frac{c\hat{f}}{2f_0}\right).
    \label{eq:theta_estimate}
\end{equation}
Zero padding is applied before the FFT to interpolate the spatial spectrum and improve the displayed peak-location granularity. The physical angular resolution remains governed by the aperture length in \Eq{eq:aperture}.

\subsection{Simulation Results}
A MATLAB simulation was implemented using the array parameters in \TAB{tab:design_parameters}. For a single target located at \(30^{\circ}\), the FFT peak gives an estimated DOA of \(30.0114^{\circ}\), as shown in \FGRu{fig:single_fft}. The estimation error is substantially below the required \(2^{\circ}\) boresight resolution.

\begin{figure}[ht!]
    \centering
    \includegraphics[width=8.8cm,height=6cm]{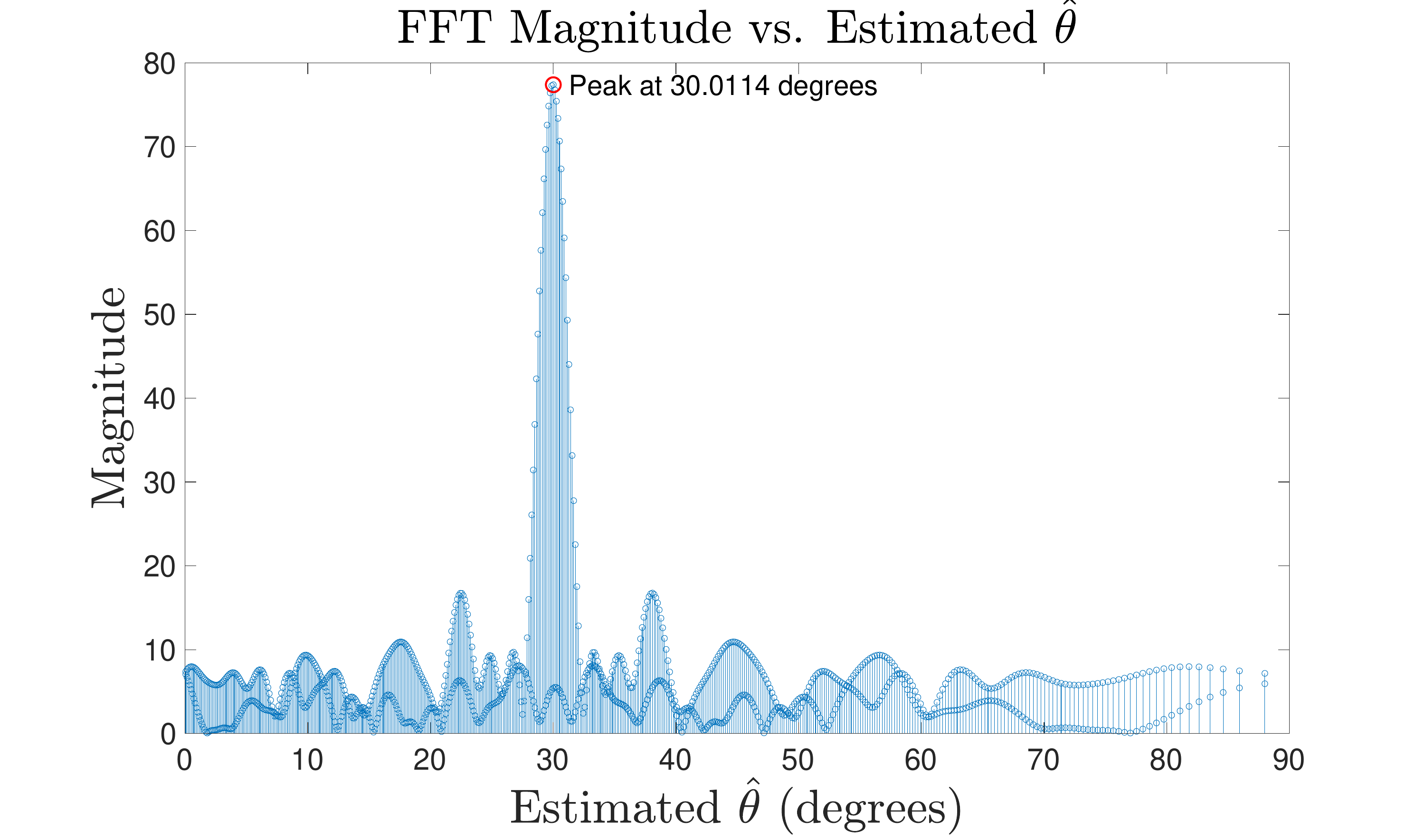}
    \caption{FFT magnitude versus estimated angle for a single target.}
    \label{fig:single_fft}
\end{figure}

The simulation was also evaluated with multiple targets. The true target angles and estimated DOAs are listed in \TAB{tab:angular_results}, and the corresponding spatial spectrum is shown in \FGRu{fig:multi_fft}. The estimates remain consistent with the specified angular-resolution constraint.

\begin{figure}[ht!]
    \centering
    \includegraphics[width=8.8cm,height=6cm]{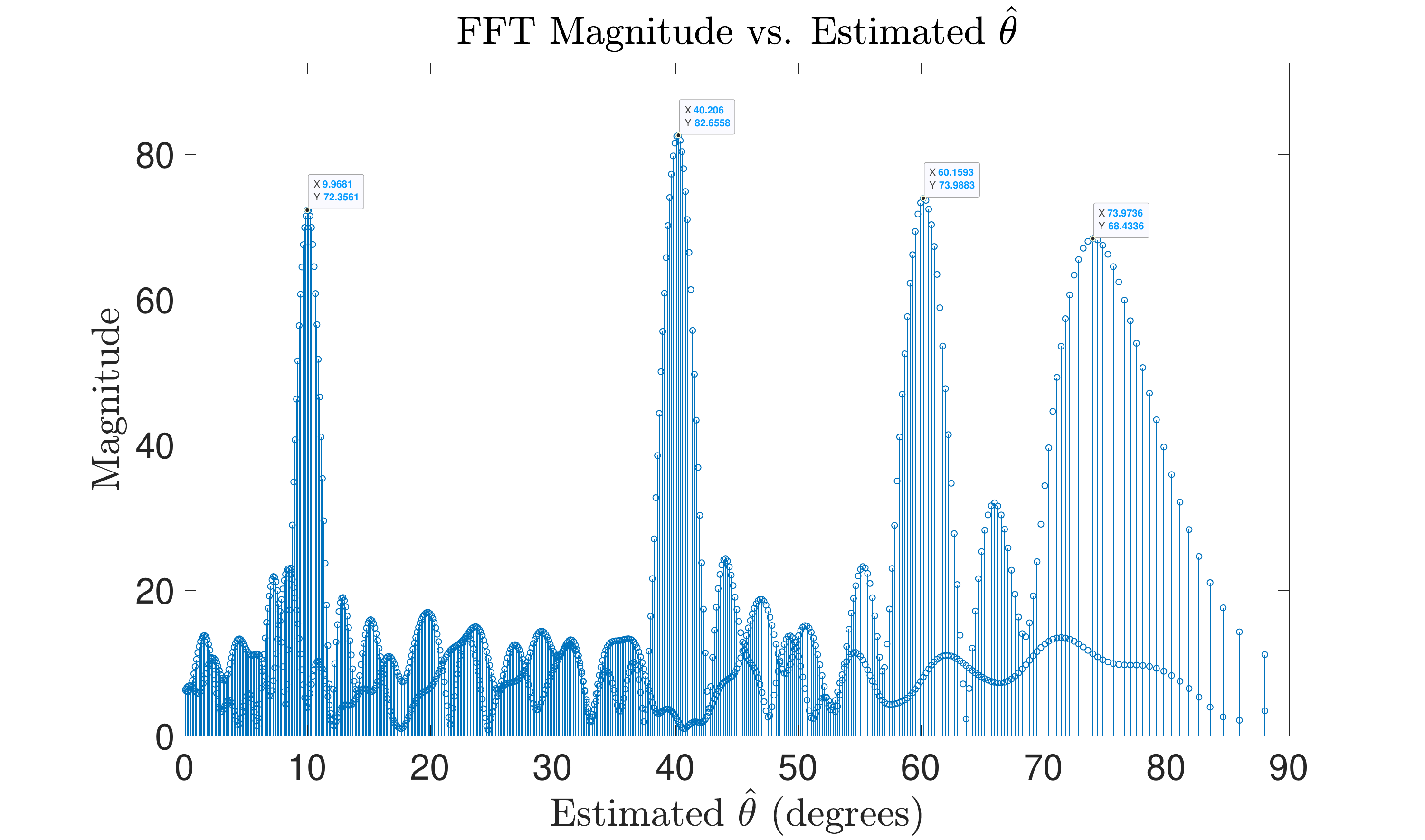}
    \caption{FFT magnitude versus estimated angle for multiple targets.}
    \label{fig:multi_fft}
\end{figure}

\begin{table}[ht!]
\caption{Angular Estimation Results}
\centering
\begin{tabular}{lccccc}
\toprule
 & \multicolumn{5}{c}{Angle} \\
\cmidrule(lr){2-6}
True DOA & \ang{10} & \ang{30} & \ang{40} & \ang{60} & \ang{75} \\
Estimated DOA & \ang{9.968} & \ang{30.011} & \ang{40.206} & \ang{60.159} & \ang{73.973} \\
\bottomrule
\end{tabular}
\label{tab:angular_results}
\end{table}

Increasing the antenna spacing decreases the spatial sampling frequency and lowers the unambiguous angular field of view. At the same time, a larger spacing increases the aperture for fixed \(N\), which narrows the main lobe. The resulting design trade-off is therefore between aliasing avoidance and aperture-driven angular resolution.

\begin{figure}[ht!]
    \centering
    \includegraphics[width=8.8cm,height=6cm]{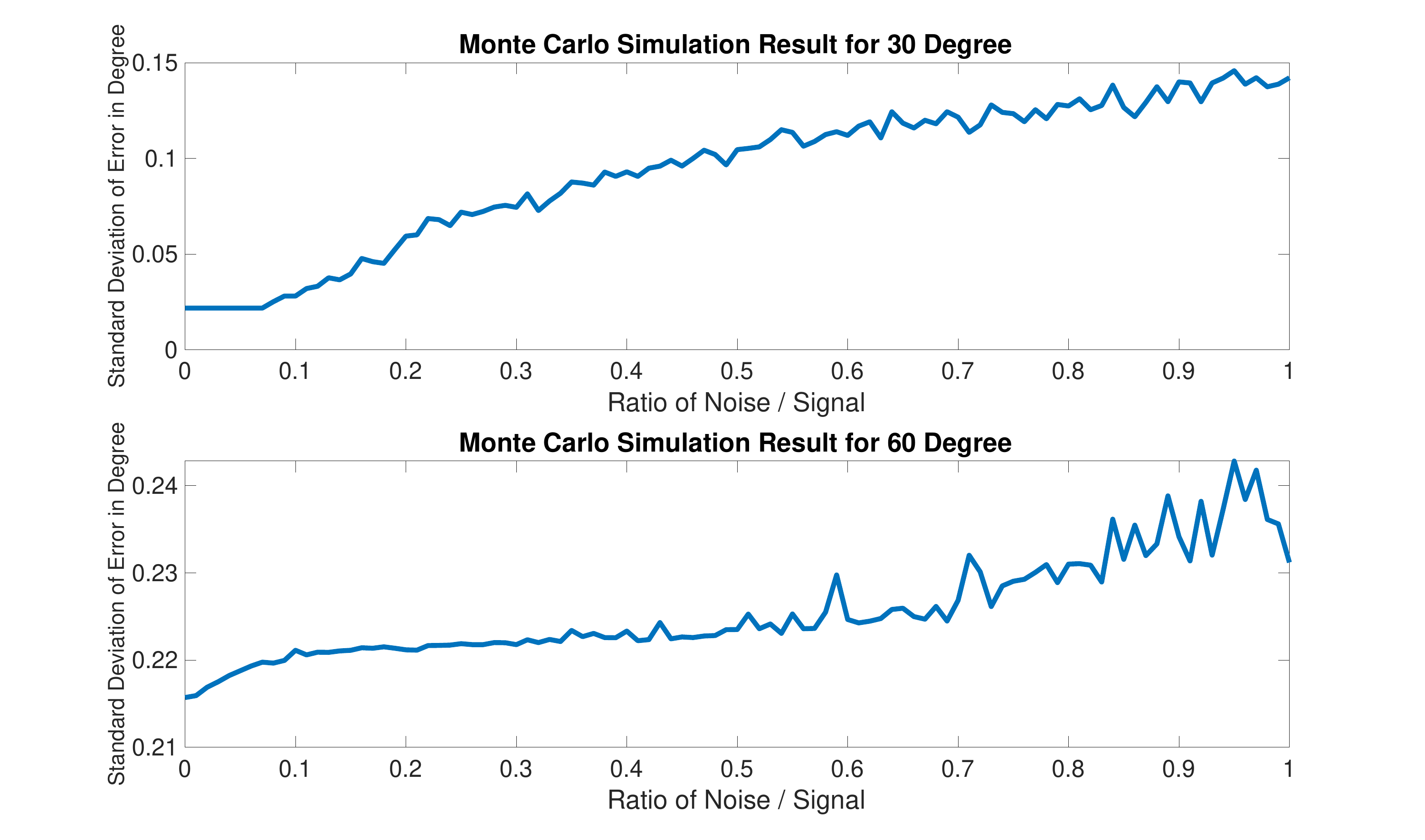}
    \caption{Standard deviation of angular error versus noise power.}
    \label{fig:noise}
\end{figure}

Additive complex Gaussian noise was introduced into the received samples, and the experiment was repeated 1000 times for targets at \(30^{\circ}\) and \(60^{\circ}\). The standard deviation of the estimation error increases with noise power, as shown in \FGRu{fig:noise}. The error at \(60^{\circ}\) is larger because the inverse-sine mapping in \Eq{eq:theta_estimate} becomes more sensitive as the angle approaches endfire. Despite this effect, the observed error remains within the target resolution over the simulated noise range.

\subsection{Decorrelating Target Response}
A decorrelating target is modeled by allowing the complex reflectivity to vary randomly between adjacent antenna measurements. This effect violates the constant-amplitude phase progression assumed by the coherent spatial FFT and therefore degrades angular estimation. The reflectivity at each element is generated from a circularly symmetric complex Gaussian random variable with unit average power. Algorithm~\ref{alg:decorrelation} summarizes the simulation procedure.

\begin{algorithm}[ht!]
\caption{Spatial FFT Simulation with Decorrelation}
\label{alg:decorrelation}
\KwIn{Target angles \(\Theta\), number of elements \(N\), FFT length \(K\)}
\KwOut{Spatial spectrum \(X_{FFT}\) and angle axis \(\theta_{axis}\)}
Initialize \(x\gets\mathbf{0}_{N\times1}\) and centered element indices \(m\)\;
\ForEach{\(\theta\in\Theta\)}{
    \(w\gets -2f_0\sin(\theta)/c\)\;
    Generate \(\alpha_m\sim\mathcal{CN}(0,1)\) for all elements\;
    \ForEach{element index \(m\)}{
        \(s_m\gets \alpha_m\exp(j2\pi w d_x m)\)\;
        Accumulate \(x_m\gets x_m+s_m\)\;
    }
}
\(X_{FFT}\gets\text{fftshift}(\text{fft}(x,K))\)\;
Map spatial-frequency bins to \(\theta_{axis}\) using \Eq{eq:theta_estimate}\;
\end{algorithm}

\begin{figure}[ht!]
    \centering
    \includegraphics[width=8.8cm,height=6cm]{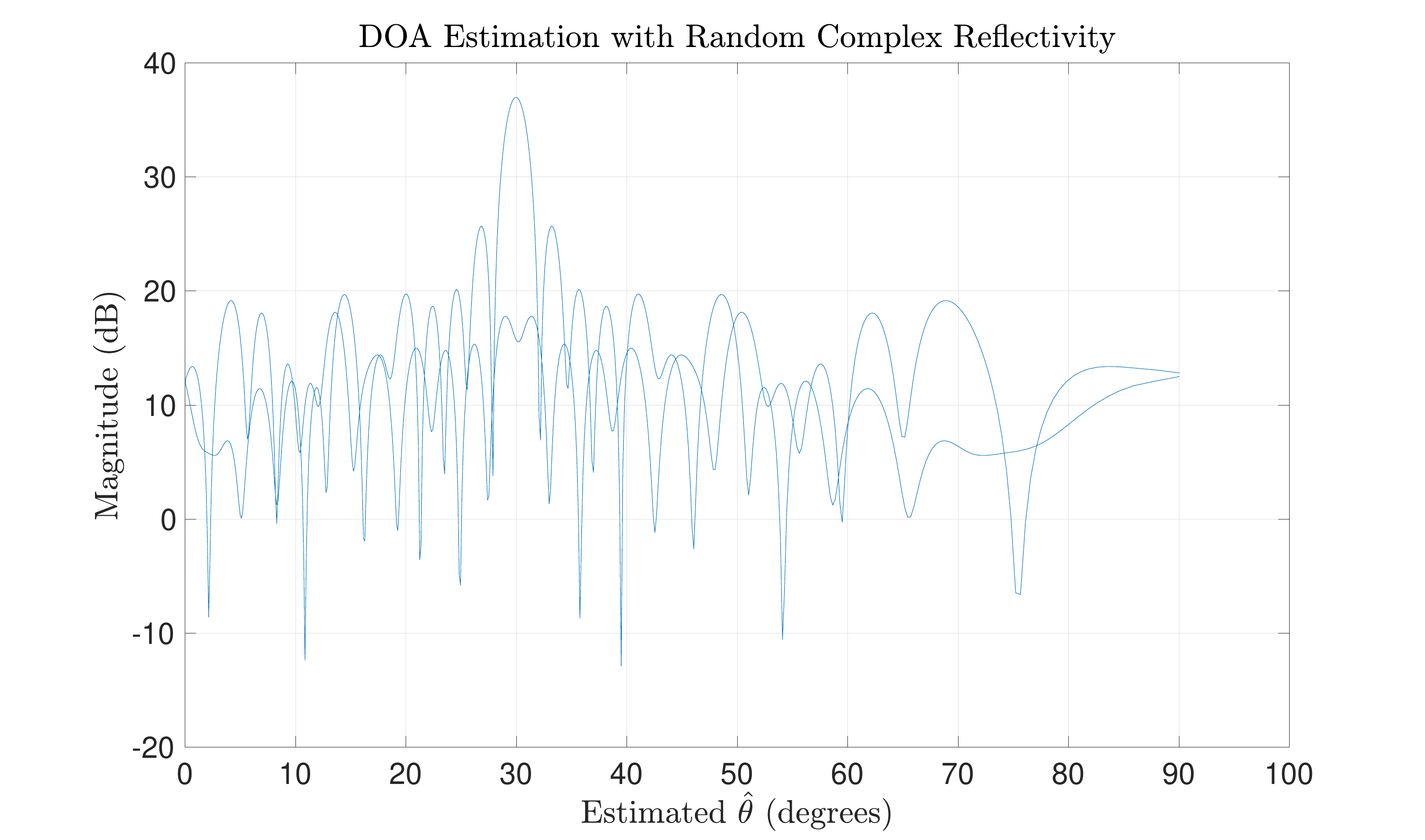}
    \caption{FFT magnitude versus estimated angle under decorrelating target reflectivity.}
    \label{fig:decorrelation_fft}
\end{figure}

The decorrelating response broadens and perturbs the spatial spectrum, as illustrated in \FGRu{fig:decorrelation_fft}. Random element-to-element reflectivity changes reduce coherent accumulation, which increases peak uncertainty and can mask weaker angular components.

\section{Two-Dimensional Position Estimation}
\subsection{Wideband Signal Model}
Narrowband spatial processing estimates angle but does not resolve range. Two-dimensional localization is obtained by replacing the sinusoidal transmission with a wideband waveform. A sinc pulse of bandwidth \(B=1\,\mathrm{GHz}\), modulated at the same carrier frequency, is used:
\begin{equation}
    g(t)=\operatorname{sinc}(Bt)e^{j2\pi f_0t}.
    \label{eq:wideband_tx}
\end{equation}
The reciprocal bandwidth gives the time resolution
\begin{equation}
    \delta t=\frac{1}{B}=1\,\mathrm{ns},
    \label{eq:time_resolution}
\end{equation}
which corresponds to the approximate range resolution
\begin{equation}
    \delta x\approx\frac{c\delta t}{2}=0.15\,\mathrm{m}.
    \label{eq:range_resolution}
\end{equation}

For a target with delay \(\tau_n\), the demodulated received signal at the \(n\)-th element is modeled as
\begin{equation}
    r_n(t)=\alpha_n\operatorname{sinc}\left(B(t-\tau_n)\right)e^{-j2\pi f_0\tau_n},
    \label{eq:wideband_rx}
\end{equation}
where \(\alpha_n\) accounts for reflectivity and propagation attenuation. The sinc peak provides range information, while the phase progression across elements at a selected range cell provides angle information. Range-angle localization is therefore achieved by first compressing the received signal in fast time and then applying spatial FFT processing across the ULA elements.

\subsection{Range-Angle Simulation Results}
A configurable MATLAB simulation was used to evaluate the wideband model. \FGRu{fig:range_fft} shows the range-domain response obtained from the sinc-modulated signal. The peak position identifies the target range, and spatial processing across the corresponding range bin estimates the target angle.

\begin{figure}[ht!]
    \centering
    \includegraphics[width=8.8cm,height=6cm]{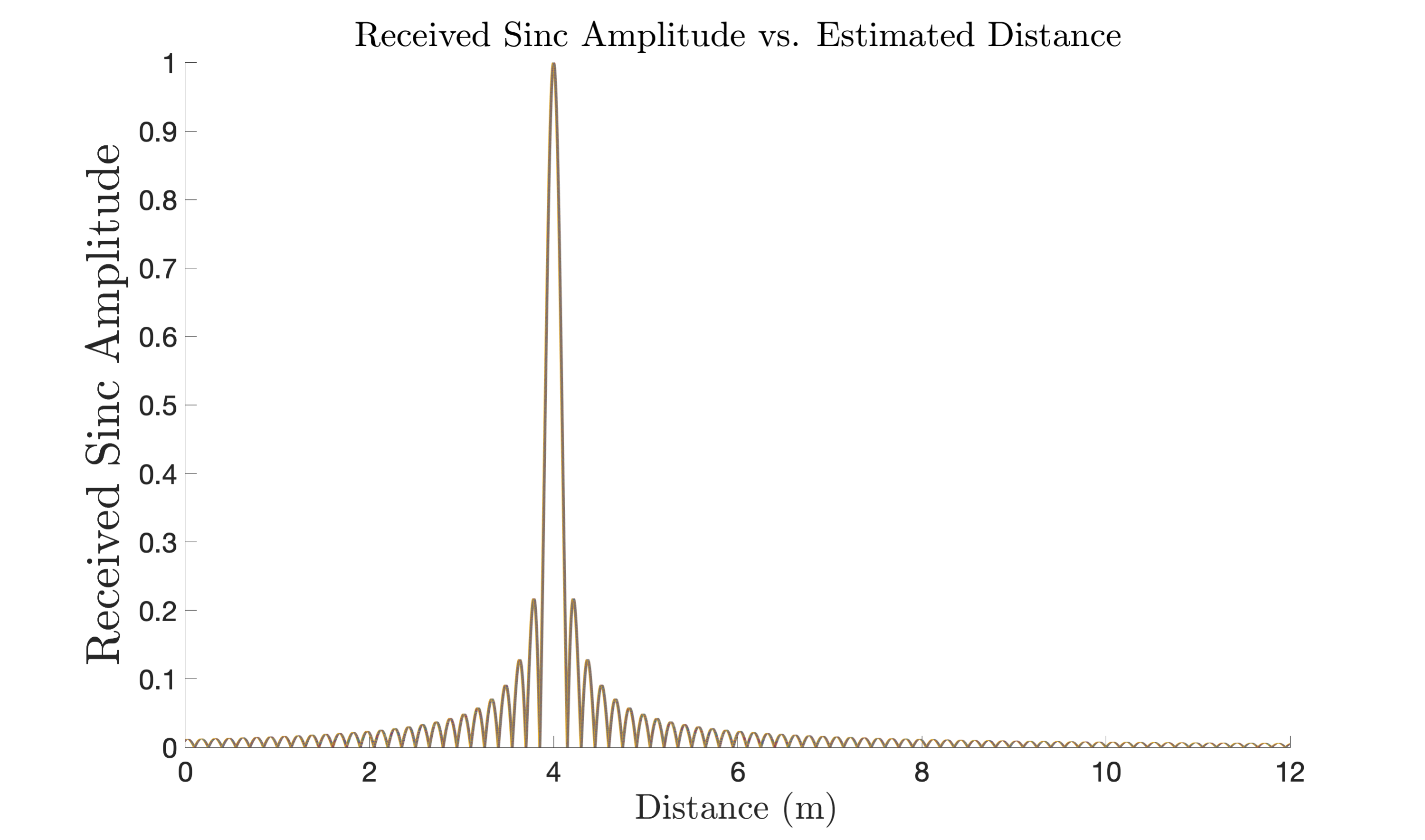}
    \caption{Magnitude response versus range after wideband compression.}
    \label{fig:range_fft}
\end{figure}

\subsubsection{Single Target at \(30^{\circ}\) and 4 m}
For a single target placed at \(30^{\circ}\) and 4 m, the range-compressed matrix is shown in \FGRu{fig:range_matrix_single}. Each row corresponds to a range sample and each column corresponds to an antenna element. The high-energy ridge follows the expected delay shift across the array. A three-dimensional visualization of the same data is shown in \FGRu{fig:range_3d_single}.

\begin{figure}[ht!]
    \centering
    \includegraphics[width=8.8cm,height=7cm]{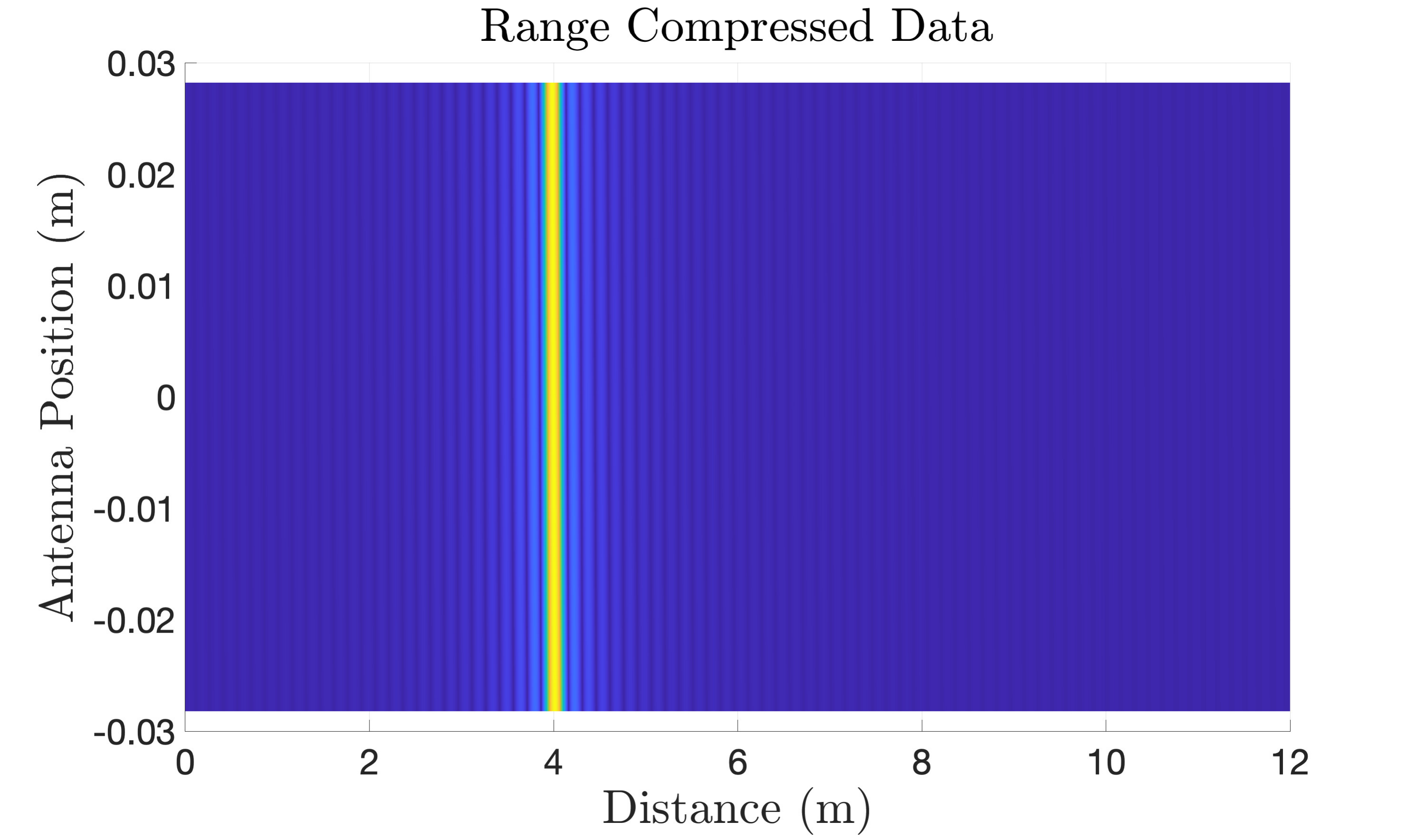}
    \caption{Range-compressed data for a single target at \(30^{\circ}\) and 4 m.}
    \label{fig:range_matrix_single}
\end{figure}

\begin{figure}[ht!]
    \centering
    \includegraphics[width=8.8cm,height=7cm]{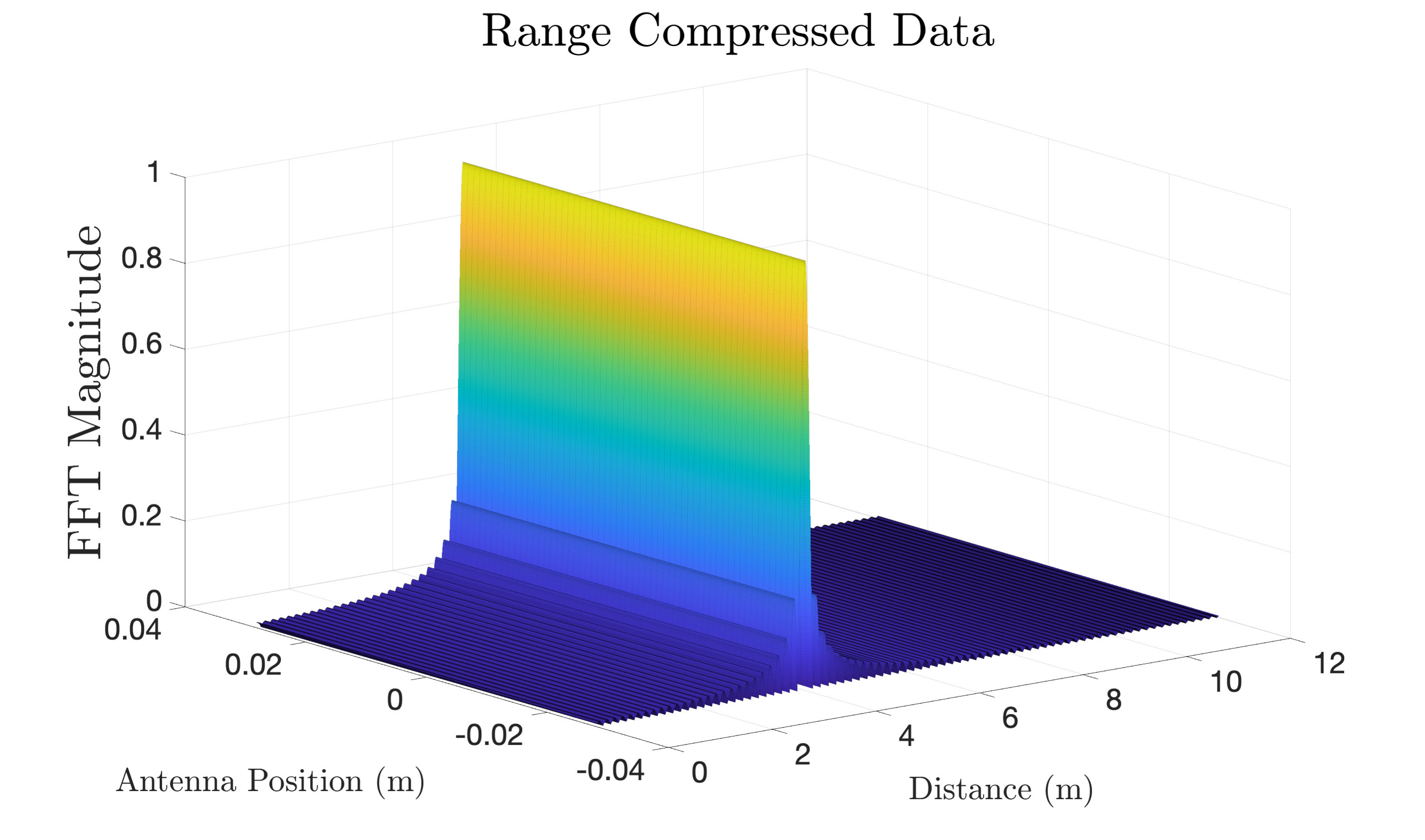}
    \caption{Three-dimensional representation of range-compressed data for a single target.}
    \label{fig:range_3d_single}
\end{figure}

\subsubsection{Effect of Antenna Spacing}
The aliasing condition remains critical for the wideband case. In the monostatic geometry, \(d_x\leq\lambda/4\) is required to prevent spatial ambiguity. When the spacing is increased beyond this limit, angular replicas appear in the FFT spectrum. \FGRu{fig:alias_fft} shows a representative case in which a target specified at \(30^{\circ}\) is estimated near \(10^{\circ}\), demonstrating spatial aliasing caused by excessive element spacing.

The figure should be read as a spatial-frequency diagnostic rather than only as an angle plot. The horizontal axis is obtained after mapping FFT bins to angle using \Eq{eq:theta_estimate}, and the vertical axis gives the normalized FFT magnitude. Because the larger spacing increases the physical aperture for a fixed number of elements, the plotted peak appears relatively narrow; however, its location is incorrect. This is the main point of the result: a sharp FFT peak does not necessarily imply a valid DOA estimate if the array has been sampled too coarsely. Once the phase advance between neighboring antennas exceeds the unambiguous interval, different values of \(\sin\theta\) generate indistinguishable phase progressions. The peak near \(10^{\circ}\) is therefore a grating-lobe ambiguity that replaces the true \(30^{\circ}\) direction with a plausible but false angle. \FGRu{fig:alias_fft} consequently emphasizes the design trade-off between resolution and field of view. Larger spacing can improve the apparent main-lobe width, but only spacings satisfying the Nyquist condition preserve a correct mapping between spatial frequency and physical angle.

\begin{figure}[ht!]
    \centering
    \includegraphics[width=8.8cm,height=5.2cm]{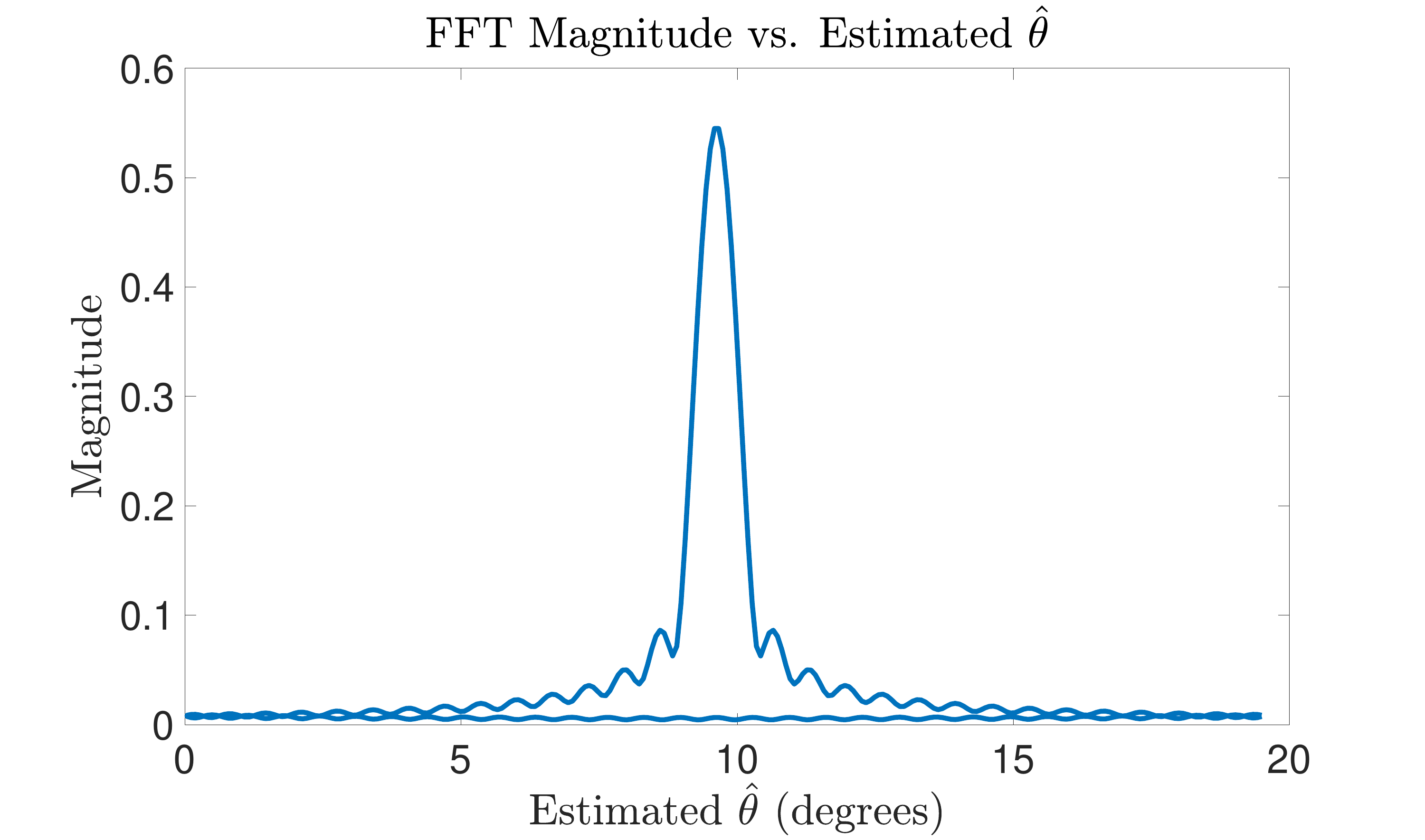}
    \caption{Aliased FFT magnitude versus estimated angle for excessive antenna spacing.}
    \label{fig:alias_fft}
\end{figure}

\subsubsection{Decorrelated Wideband Target}
Target decorrelation also affects the range-angle representation. \FGRu{fig:decorrelated_range} shows the range-compressed data under decorrelated reflectivity, and \FGRu{fig:decorrelated_3d} shows the corresponding magnitude and range-domain representation. Compared with the coherent case, the decorrelated target produces stronger variability across antenna elements and a less stable spatial signature, which reduces the reliability of the angular peak.

\begin{figure}[ht!]
    \centering
    \includegraphics[width=8.8cm,height=7.5cm]{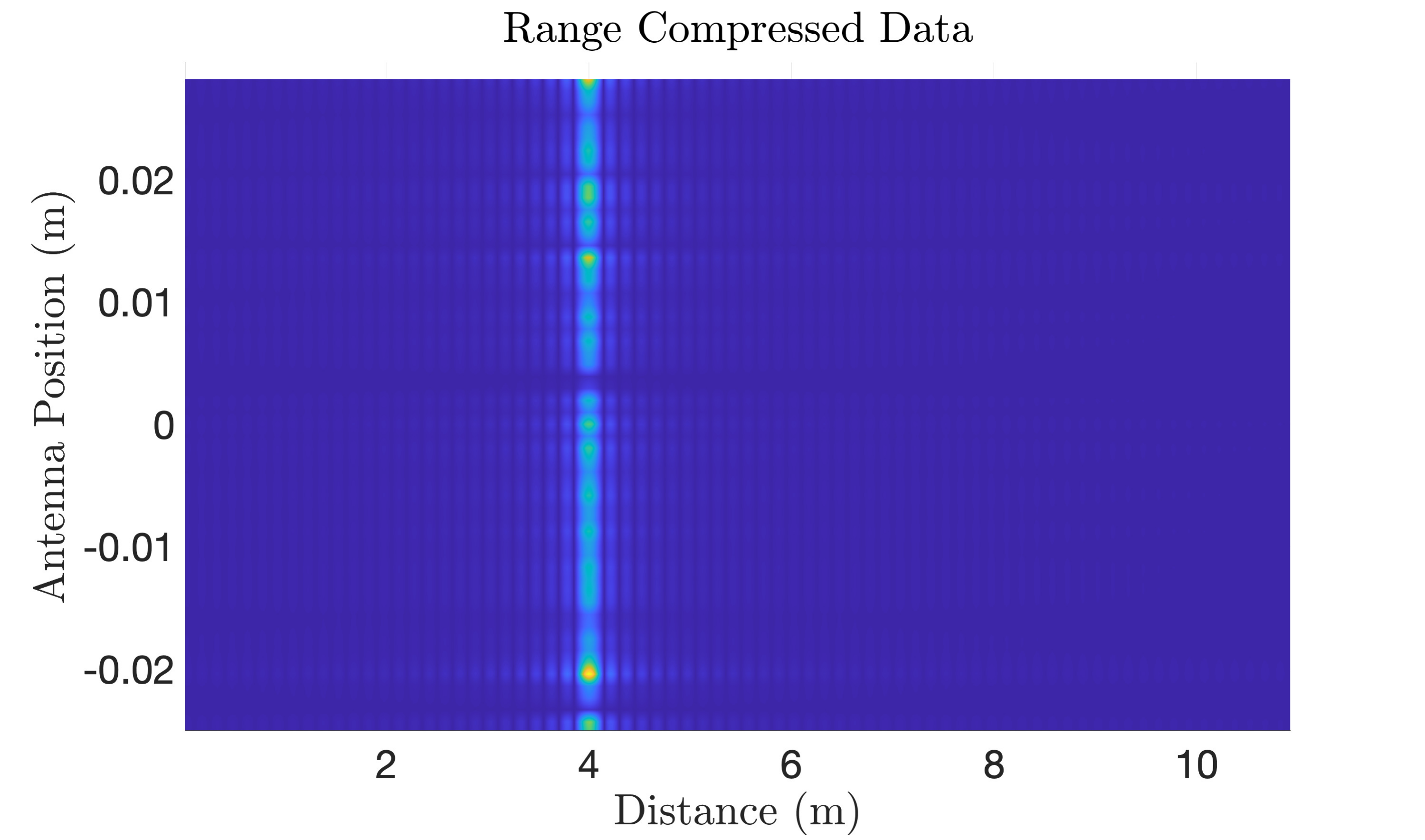}
    \caption{Range-compressed data for a decorrelated target.}
    \label{fig:decorrelated_range}
\end{figure}

In \FGRu{fig:decorrelated_range}, the target energy is still concentrated around the expected range because range compression is performed independently on the fast-time signal at each antenna. However, the response is no longer smooth along the antenna-position axis. Adjacent columns exhibit random bright and weak regions, which indicates that the complex reflectivity is changing from element to element instead of following a single deterministic steering vector. This visual pattern explains why the range estimate remains recognizable while the angle estimate becomes less reliable. The range peak depends mainly on delay alignment, whereas the DOA FFT depends on coherent phase accumulation across the whole aperture.

\FGRu{fig:decorrelated_3d} presents the same effect in three dimensions. Instead of the clean, continuous ridge observed for the coherent single-target case, the surface contains several uneven spikes and local maxima around the target range. These spikes correspond to antenna samples that happen to add constructively after random reflectivity weighting, while neighboring samples may partially cancel. As a result, the spatial FFT receives an aperture response with amplitude and phase discontinuities. The dominant angular lobe is broadened and perturbed, and weaker false components may appear depending on the particular random realization. The two decorrelation figures therefore provide a useful physical interpretation of the numerical result: target coherence is not only a propagation assumption, but also a practical requirement for obtaining a stable and high-contrast angular spectrum.

\begin{figure}[ht!]
    \centering
    \includegraphics[width=8.8cm,height=7.5cm]{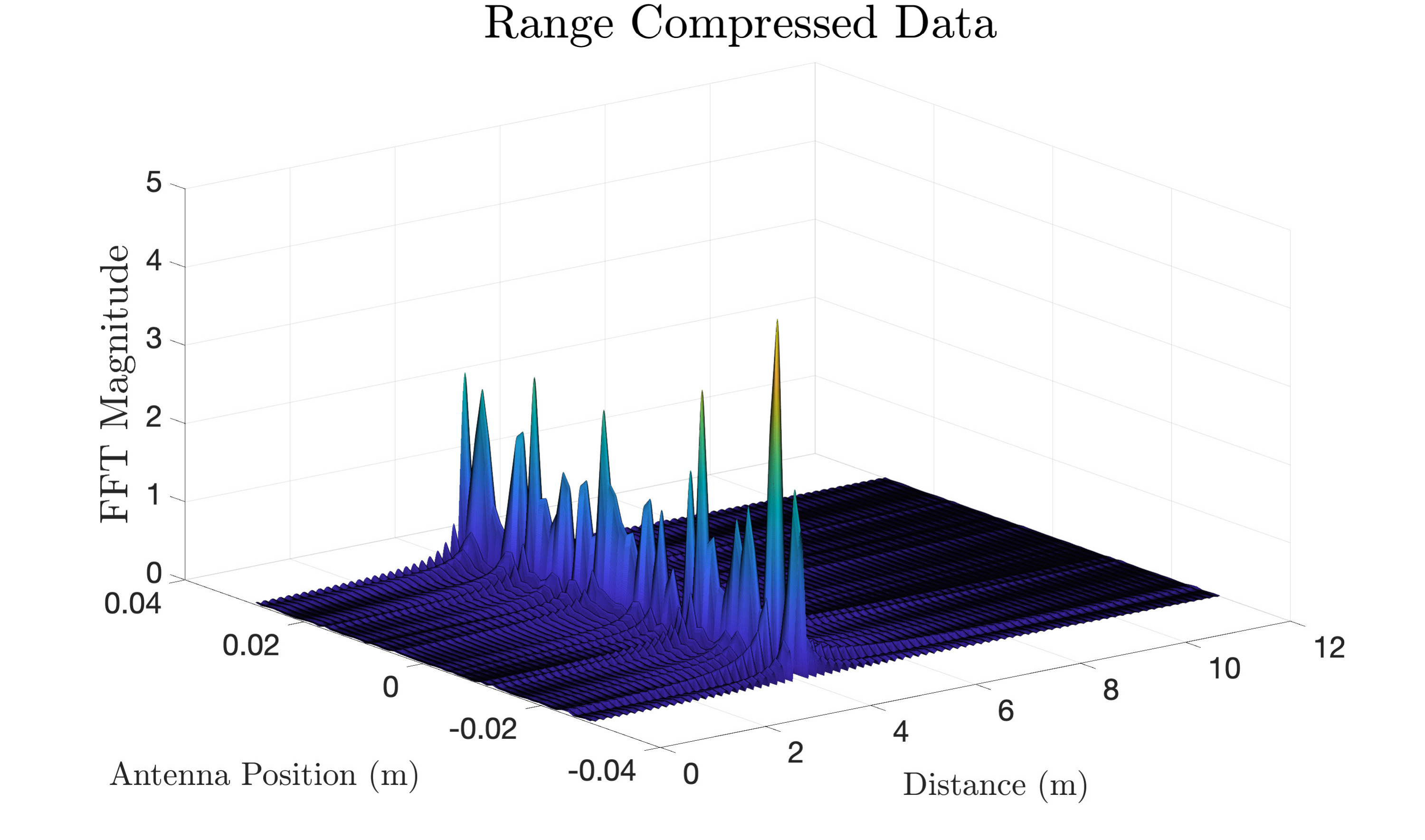}
    \caption{Magnitude and range-domain representation for the decorrelated target.}
    \label{fig:decorrelated_3d}
\end{figure}

\section{Conclusion}
\enlargethispage{2\baselineskip}
An FFT-based localization framework for a 77 GHz monostatic ULA has been formulated and evaluated. The narrowband derivation shows that DOA estimation can be treated as spatial-frequency estimation across the array. The spatial sampling condition leads to an element spacing of \(d_x=c/(4f_0)=0.97276\) mm, and the aperture constraint for \(2^{\circ}\) boresight resolution leads to a minimum of 58 elements. Simulation results confirm accurate estimation for a single target at \(30^{\circ}\) and for multiple targets distributed between \(10^{\circ}\) and \(75^{\circ}\). The single-target and multi-target spectra show that the FFT method produces clear peaks when the received samples follow the expected steering-vector model. The noise experiment then clarifies the robustness limit of this behavior: as the complex noise power increases, the peak location fluctuates more strongly, especially at larger angles where the inverse-sine conversion from spatial frequency to DOA is more sensitive. The decorrelation experiment gives a complementary limitation by showing that random element-to-element reflectivity changes reduce coherent summation, broaden the FFT response, and can obscure weaker angular components even when the nominal target range is still visible.

The wideband extension demonstrates that range estimation can be incorporated by using a 1 GHz sinc-modulated waveform. The corresponding time resolution of 1 ns gives an approximate range resolution of 0.15 m, enabling range-angle localization through range compression followed by spatial FFT processing. The range-domain and range-matrix figures show how the wideband waveform separates delay from direction: first, the compressed pulse identifies the distance cell; second, the samples across the ULA at that cell provide the spatial phase progression needed for angle estimation. The coherent single-target plots contain a compact and smooth response around 4 m, while the aliasing and decorrelation plots show two different failure modes. Excessive element spacing gives a precise-looking but wrong angular peak because the spatial samples violate the alias-free condition. Decorrelated reflectivity preserves the approximate range location but breaks the smooth aperture signature required for a stable DOA estimate. Together, the final figures connect the visual plots to the design rules: the coherent case shows the desired response, the aliasing case shows a false but sharp angle, and the decorrelated case shows coherence loss. Thus, visual sharpness must always be interpreted together with the sampling and coherence assumptions.

Overall, the results highlight that reliable ULA-based positioning requires simultaneous control of aperture length, element spacing, signal bandwidth, and target coherence. Aperture length determines the ability to separate nearby angles, element spacing determines whether the estimated spatial frequency is unambiguous, bandwidth determines the range resolution, and coherence determines whether the array samples add constructively in the FFT beamformer. The proposed design satisfies these constraints for the simulated 77 GHz scenario and provides a compact demonstration of how narrowband DOA estimation can be extended into practical range-angle localization.

\balance
\bibliographystyle{IEEEtran}
\bibliography{references}

\end{document}